\documentclass[epj,final]{svjour}

\usepackage{amsmath}
\usepackage{amssymb}   
\usepackage{graphicx}

\DeclareGraphicsExtensions{.eps}
\newlength{\figwidth}

\begin{document}

\title{Effect of flux-dependent Friedel oscillations 
upon the effective transmission of an interacting nano-system}

\titlerunning{Effective transmission of an interacting nano-system}

\author{%
Axel Freyn 
\and 
Jean-Louis Pichard}
\authorrunning{A.\ Freyn and J.-L.\ Pichard}

\institute{%
Service de Physique de l'Etat Condens{\'e} (CNRS URA 2464),
DSM/DRECAM/SPEC, \\ 
CEA Saclay, 91191 Gif sur Yvette Cedex, France
}

\date{\today}

\abstract{%
We consider a nano-system connected to measurement probes via 
non interacting leads. When the electrons interact inside the 
nano-system, the coefficient $|t_s(E_F)|^2$ describing its 
effective transmission at the Fermi energy $E_F$ ceases to be 
local. This effect of electron-electron interactions upon 
$|t_s(E_F)|^2$ is studied using a one dimensional model of spinless 
fermions and the Hartree-Fock approximation. The non locality of 
$|t_s(E_F)|^2$ is due to the coupling between the Hartree and Fock 
corrections inside the nano-system and the scatterers outside the 
nano-system via long range Friedel oscillations. Using this phenomenon, 
one can vary $|t_s(E_F)|^2$ by an Aharonov-Bohm flux threading a ring 
which is attached to one lead at a distance $L_c$ from the nano-system. 
For small distances $L_c$, the variation of the quantum conductance 
induced by this non local effect can exceed $0.1 (e^2/h)$.
}

\maketitle

%%%%%%%%%%%%%%%%% INTRODUCTION %%%%%%%%%%%%%%%%%%%%%%

\section{Introduction}

% 
% *********** Summary of the previous works *******
%               
%
%
%
%
 In the scattering approach \cite{Landauer,Buttiker,Imry} to quantum 
transport, the measure of the quantum conductance $g$ of a nano-system 
requires incoherent electron reservoirs at a temperature $T$ and metallic 
contacts (non interacting leads). For a two-probe measurement and one 
dimensional (1d) leads, $g$ (in units of $e^2/h$ for spin polarized 
electrons) is given in the limit $T \rightarrow 0$ by the probability 
$|t_s(E_F)|^2$ that an electron emitted from one reservoir at the Fermi 
energy $E_F$ can be transmitted to the other reservoir through the 
nano-system and its attached leads. If the electron-electron interactions 
are negligible inside the nano-system, $|t_s(E_F)|^2$ is a local quantity 
which is independent of other scatterers that the attached leads can have. 
If the electrons interact inside the nano-system, the definition of 
$|t_s(E_F)|^2$ becomes much more subtle, since the nano-system is no 
longer a one body scatterer, but a many-body scatterer. Fortunately, a 
many body scatterer with two attached non interacting leads behaves 
when $T \rightarrow 0$ as an effective one body scatterer with interaction 
dependent parameters, and its effective one body transmission determines 
its quantum conductance. 

A numerical proof of this statement is given in Ref. \cite{Molina1}, based 
on the study of a ring made of a 1d auxiliary lead embedding a nano-system. 
The electrons were assumed without interaction unless being inside 
the nano-system. The persistent current $I$ was numerically calculated as 
a function of the flux $\Phi$ piercing the ring. The values of $I(\Phi)$ 
were accurately determined using the DMRG algorithm \cite{DMRG1,DMRG2} for 
an auxiliary lead of length $L_L$, and extrapolated to their limits as 
$L_L \rightarrow \infty$. The extrapolated values of $I(\Phi)$ calculated 
when an interaction of strength $U$ acts inside the nano-system were shown 
to be identical to those given by a one body scatterer with an interaction 
dependent transmission coefficient $|t_s(E_F,U)|^2$. The embedding method 
\cite{Molina1,Molina2,Molina3,Gogolin,Mila,Sushkov,Meden,Rejec} 
consists in obtaining $|t_s(E_F,U)|^2$ from the extrapolated values of 
$I(\Phi)$. 

However, an important difference between the many body problem 
and the one body problem is pointed out in Ref. \cite{MWP}. Studying  
two identical interacting nano-systems in series by the embedding 
method, one finds that the value of $|t_s(E_F,U)|^2$ characterizing 
the transmission of the first nano-system is modified by the presence 
of the second nano-system. $L_c$ being the length of the ideal wire 
coupling the two nano-systems, the correction induced by the second 
nano-system upon the effective transmission $|t_s(E_F,U)|^2$  of the 
first nano-system decays as $1/L_c$, with oscillations of period equal 
to half the Fermi wave length $\lambda_F/2$. This decay characterizes 
also the Friedel oscillations of the electron density induced by a 
scatterer inside a 1d non interacting electron gas. The presence of 
this correction to $|t_s(E_F,U)|^2$ shows us that this is not the 
interacting nano-system itself, but the nano-system with its contacts 
(attached leads and embedded scatterers) which is described by 
$|t_s(E_F,U)|^2$. The decay of this correction suggests that it 
is a consequence of the Friedel oscillations of the conduction electrons 
inside the coupling wire, which are caused by the two nano-systems 
in series. 

If the DMRG studies can give accurate results, the Hartree-Fock (HF) 
approximation has the merit to give a simple explanation for this non 
local transmission. This was done in Ref. \cite{AFP}, considering the 
Hartree and Fock corrections due to a local interaction inside a 
nano-system. In a tight-binding model, the Hartree corrections modify 
the site potentials seen by a transmitted electron, while the exchange 
terms give corrections to the hopping integrals. These HF corrections 
probe energy scales below $E_F$ and length scales larger than the size 
of the nano-system inside which the electrons interact. Putting a second 
scatterer at a distance $L_c$ from the interacting nano-system induces 
Friedel oscillations of the electron density inside the nano-system, which 
change the nano-system HF corrections. This means that the effective 
scattering properties of interacting nano-systems in series are coupled 
between themselves, exactly as are coupled magnetic moments by the RKKY 
interactions \cite{RK,Y,VV,BF}. 

\begin{figure}
\centerline{\includegraphics[width=\linewidth]{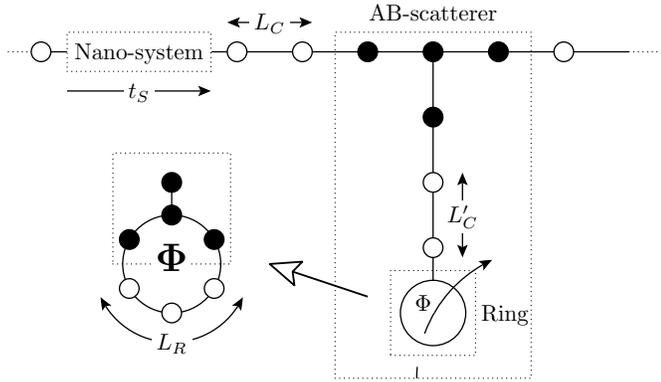}}
\caption{Considered set-up made of a many body scatterer of effective 
transmission $|t_s|^2$ with two semi-infinite 1d leads: Polarized 
electrons interact only inside the nano-system (two sites with inter-site 
repulsion $U$, hopping term $t_d$ and site potentials (gate voltage) 
$V_G$). A ring is attached at a distance $L_c$ from the nano-system. 
The nano-system is described in more details in Fig. \ref{fig4}.
} 
\label{fig1} 
\end{figure} 
\begin{figure}
\centerline{\includegraphics[width=\linewidth]{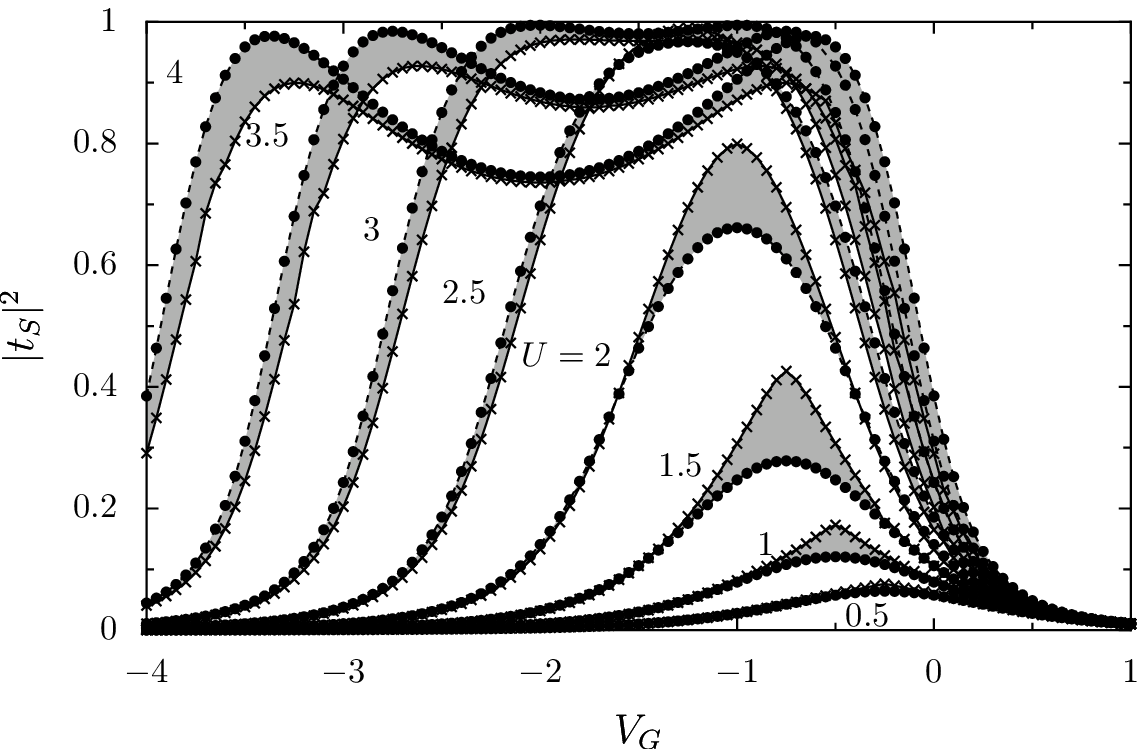}}
\caption{
Effective transmission $|t_s|^2$ as a function of the gate voltage $V_G$, 
at half filling (Fermi momentum $k_F=\pi/2$) and for a nano-system hopping 
term $t_d=0.1$. The AB-scatterer with its attached ring 
($L'_c=4, L_R=7$) is at $L_c=2$ sites from the nano-system. The interaction 
strength $U$ is indicated in the figure. A flux $\Phi=0$ ($\bullet$) or 
$\Phi=\Phi_0/2$ (x) threads the ring. The grey areas underline the effect 
of $\Phi$ upon $|t_s|^2$.
}
\label{fig2}
\end{figure}
\begin{figure}
\centerline{\includegraphics[width=\linewidth]{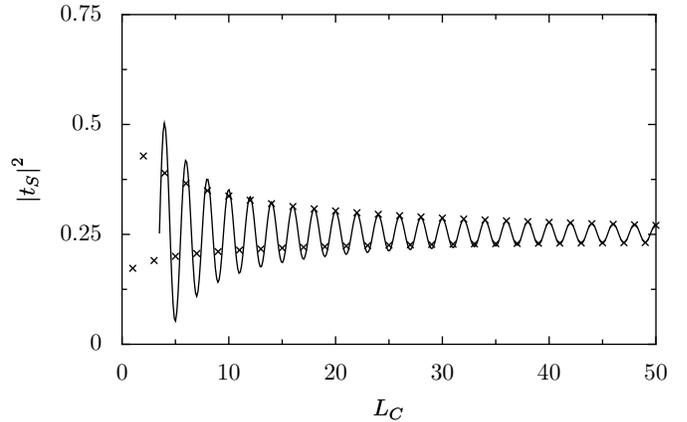}}
\caption{
$|t_s|^2$ as a function of the length $L_c$ between the 
nano-system and the ring when $\Phi=0$: The effect of the ring 
upon $|t_s|^2$ (see Fig. \ref{fig2}) decays as Friedel oscillations. 
HF results (x) and fit $0.2522 + \cos(\pi L_C)/L_C$ (solid line), 
calculated for $V_G=-0.75$ and $U=1.5$ ($L'_c=4$, $L_R=7$, $k_F=\pi/2$, 
$t_d=0.1$).
}
\label{fig3}
\end{figure}

 This effect was studied in a previous letter \cite{FP-PRL}, assuming a 
set-up which can be convenient for an experimental check of the theory: 
an infinite 1d tight-binding model of spin polarized electrons (spinless 
fermions), embedding two scatterers separated by $L_c$ sites, as sketched in 
Fig. \ref{fig1}. The first scatterer is the nano-system inside which the 
electrons interact, while the second contains an attached ring. Hereafter, 
we refer to the second scatterer with its attached ring as the AB-scatterer, 
since an Aharonov-Bohm (AB) flux $\Phi$ can pierce the ring, its variations 
inducing periodic AB-oscillations of the electron density inside the 
nano-system. This yields flux dependent HF corrections for the nano-system, 
and hence AB-oscillations of its effective transmission $|t_s|^2$. This 
non local effect upon $|t_s|^2$ induced by a ring attached at $L_c$ sites 
from the nano-system is a pure many body effect which was the subject of 
Ref. \cite{FP-PRL}. In this longer paper, a detailed derivation of the 
results summarized in Ref. \cite{FP-PRL} is given, with new analytical and 
numerical results showing how one can make this effect very large. 

 Using the set-up sketched in Fig. \ref{fig1}, the non local effect upon 
$|t_s|^2$ is illustrated in Fig. \ref{fig2}. The nano-system effective 
transmission $|t_s|^2$ is given as a function of a gate voltage $V_G$ 
applied upon the nano-system, at half filling (Fermi momentum $k_F=\pi/2$). 
For each strength $U$ of a nearest neighbor repulsion acting inside the 
nano-system, two curves give $|t_s|^2$ as a function of $V_G$ when 
the ring is attached near the nano-system ($L_c=2$). The first curve 
(full circle) has been calculated when there is no flux $\Phi$ threading 
the ring, while the second one (cross) gives $|t_s|^2$ when half a flux 
quantum $\Phi_0/2$ threads the ring. If $U=0$, the two curves are identical. 
The effect of $U$ consists in changing the shape of the curves 
$|t_s(V_G)|^2$, and in making a difference underlined by grey areas 
between the cases where $\Phi=0$ and $\Phi=\Phi_0/2$. Around certain values 
of $V_G$, the effect of $\Phi$ upon $|t_s|^2$ is of order $0.2$ for a 
transmission $|t_s|^2 \leq 1$. This means that one can make the effect 
huge if $L_c$ is small, for well chosen values of the nano-system 
parameters. Fig. \ref{fig3} shows how the effect of the ring upon $|t_s|^2$ 
decays as $L_c$ increases. One can see the $\cos (2 k_F L_c)/L_c$ 
asymptotic decay with even-odd oscillations characteristic of Friedel 
oscillations at half-filling. 

  In this paper, we explain the origin of the non local effects upon 
$|t_s|^2$ shown in Fig. \ref{fig2}. The paper is divided as follows. In 
section \ref{section1}, the nano-system Hamiltonian is defined and the HF 
equations are given when it is embedded between two semi-infinite ideal 
leads. One gets two coupled equations which have to be solved 
self-consistently. The two equations are explicitly derived when the 
nano-system is not in series with another scatterer. A numerical method 
for having the HF parameters is then defined, which allows us to recover 
the results of the analytical derivations and to estimate its convergence 
when the size of the leads increases. In section \ref{section2}, a simple 
limit where the HF parameters take trivial values is studied. In 
this limit, one can easily calculate the transmission $|t_s|^2$ as a 
function of $V_G$ at a given strength $U$ of the interaction, and explain 
the shape of the curves $|t_s(V_G)|^2$ shown in Fig. \ref{fig2}. 
Unfortunately, this limit is also the limit where the non local effect 
upon $|t_s|^2$  is negligible. For having large effects, one needs to 
be in the opposite limit. In section \ref{section3}, the scatterer with 
the attached ring (AB-scatterer) is defined. Its scattering properties 
are calculated for each energy $E\leq E_F$. In section \ref{section4}, 
the oscillations induced by the nano-system and by the AB-scatterer in the 
leads are studied separately, illustrating the phenomena responsible for 
the non locality of $|t_s|^2$. In section \ref{section5}, one considers the 
interacting nano-system in series with the AB-scatterer, and we study the 
role of the gate potential $V_G$, the Fermi momentum $k_F$ and the hopping 
term $t_d$ upon the flux dependence of $|t_s|^2$. In section \ref{section6}, 
the implications of the non locality of  $|t_s|^2$ upon the total 
quantum conductance $g_T$ are studied, when the nano-system is in series 
with the AB-scatterer between two measurement probes. We give a short summary 
in section \ref{section7}, underlining the possible relevance of the many 
body effect described in this work for the theory of experiments 
imaging coherent electron flow from a quantum point contact in a two 
dimensional electron gas. 

\section{Hartree-Fock description of an interacting nano-system 
with non interacting leads} 
\label{section1}

 We consider a one dimensional tight-binding model of spin polarized 
electrons (spinless fermions), where the particles do not interact, 
unless they occupy two nearest neighbor sites ($0$ and $1$), which 
costs an interaction energy $U$. The two sites $0$ and $1$, with 
potentials $V_0=V_1=V_G$, a repulsion $U$ and an hopping term $t_d$ 
define the nano-system. We assume that the potential $V_G$ can be 
varied by a gate. The nano-system Hamiltonian reads 
\begin{equation}
H_{s} = - t_d (c^\dagger_0 c^{\phantom{\dagger}}_{1} + h.c.)     
+ V_G (n_{1}+n_{0}) + \ U n_{1} n_{0} \, .
\label{hamiltonian-system}
\end{equation}
$c^{\phantom{\dagger}}_p$ ($c^\dagger_p$) is the annihilation (creation) 
operator at site $p$, and $n_p = c^\dagger_p c^{\phantom{\dagger}}_p$.
The left (L) and right (R) leads are described by two Hamiltonians 
\begin{equation}
H_{lead}^{L,R} = -\sum_{p} t_h 
(c^\dagger_{p-1} c^{\phantom{\dagger}}_{p} + h.c.),     
\label{hamiltonian-lead}
\end{equation}
where $p$ runs from $-\infty$ to $-1$ ($3$ to $\infty$) for the left (right) 
lead. The hopping amplitude in the leads $t_h=1$ sets the energy 
scale, the conduction band corresponding to energies $-2 <E=-2\cos k <2$ 
($k$ real). The two leads and the nano-system are coupled by 
\begin{equation}
H_{coupling}^{L,R} = - t_c 
(c^\dagger_{p-1} c^{\phantom{\dagger}}_{p} + h.c.)     
\label{hamiltonian-coupling}
\end{equation}
with $p=2$ ($0$) for the coupling with the right (left) lead. 
The Hamiltonian 
\begin{equation}
H=H_{s}+ \sum_{J=L,R} (H_{lead}^{J}+H_{coupling}^{J})
\label{hamiltonian-total}
\end{equation} 
defines the interacting nano-system coupled with two non interacting 1d 
semi-infinite leads.

\begin{figure}
\centerline{\includegraphics[width=\linewidth]{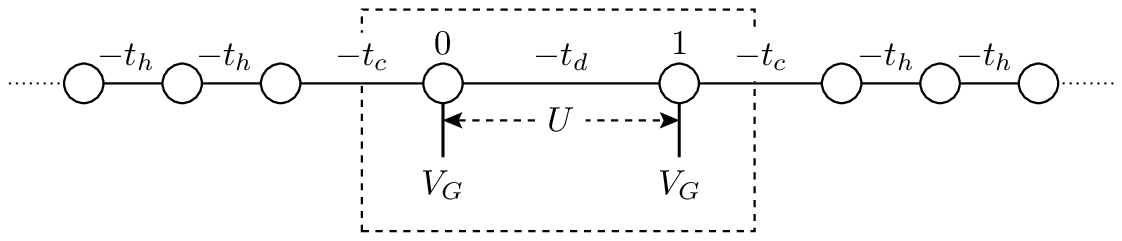}}
\caption{Nano-system with two semi-infinite 1d leads: Spin polarized 
electrons interact only inside the nano-system (sites $0$ and $1$) 
with inter-site repulsion $U$ and site-potentials $V_0=V_1=V_G$. 
The strength of the hopping terms is $t_d$ inside the nano-system, 
$t_c$ between the nano-system and the leads, and $t_h=1$ in the leads.}
\label{fig4} 
\end{figure}

  In the HF approximation, one takes for the ground state a 
Slater determinant of one-body wave-functions $\psi_{\alpha}(p)$ of 
energy $E_{\alpha} < E_F=-2 \cos k_F$. The $\psi_{\alpha}(p)$ are the 
eigenfunctions of the Hamiltonian $H$ (Eq. \ref{hamiltonian-total}), 
where the nano-system is described by an effective one body Hamiltonian  
\begin{equation}
H_{s}^{HF} = - v (c^\dagger_0 c^{\phantom{\dagger}}_{1} + h.c.)     
+ V (n_{1}+n_{0}) \, 
\label{hamiltonian-system-HF}
\end{equation}
instead of $H_s$. $H_{s}^{HF}$ does not contain the two body term 
$U n_1 n_0$ of $H_s$ (Eq. \ref{hamiltonian-system}), but a renormalized 
hopping term $v$ (instead of $t_d$), and a renormalized gate potential 
$V$ (instead of $V_G$). The form of $H_{s}^{HF}$ results from the 
nearest neighbor repulsion acting only between the sites $0$ and $1$, 
such that the exchange correction modifies only the strength of the 
hopping term $t_d$ coupling those two sites, while the site-potentials 
$V_0$ and $V_1$ have two identical Hartree corrections, because of the 
reflection symmetry $p-1/2 \rightarrow -p+1/2$. 

For the HF calculations, we proceed in three steps. All  
wave-functions $\psi_{\alpha}(p)$ of energy $E_{\alpha}\leq E_F$ 
are calculated for arbitrary values of $v$ and $V$. Then, 
the expectation values
\begin{equation}
\begin{aligned}
\left \langle c^\dagger_0 c^{\phantom{\dagger}}_{1} (v,V)
\right \rangle = & \sum_{E_{\alpha}<E_F}\psi^*_{\alpha}(0)\psi_{\alpha}(1) \\ 
\left\langle c^\dagger_0 c^{\phantom{\dagger}}_{0} (v,V)
\right \rangle = & \sum_{E_{\alpha}<E_F}\psi^*_{\alpha}(0)\psi_{\alpha}(0) 
\end{aligned}
\end{equation}
are evaluated, either analytically or numerically. Eventually, the values 
of the two HF parameters $v$ and $V$ are adjusted till they converge towards 
the two self-consistent values which satisfy the coupled integral equations: 
\begin{equation}
\begin{aligned}
v =& t_d+U \left \langle c^\dagger_0 c^{\phantom{\dagger}}_{1} (v,V)
\right \rangle   \\ 
V= & V_G+U \left\langle c^\dagger_0 c^{\phantom{\dagger}}_{0} (v,V)
\right \rangle
\label{HF}
\end{aligned}
\end{equation}

Once the self-consistent values of $v$ and $V$ are numerically 
obtained from Eqs. (\ref{HF}), the effective transmission amplitude 
$t_{s}$ of the nano-system at an energy $E_F=-2 \cos k_F$ reads:  
\begin{equation}
t_{s}(U)=\frac{v(1-e^{-2ik_F})}
{v^2-e^{-2ik_F}-2V e^{-ik_F} - V^2}.
\label{transmission}
\end{equation}

\subsection{Analytical form of the HF-equations}
\label{AFHFE}

 For the nano-system with two semi-infinite leads, the two first steps 
can be done analytically, while the last step requires to numerically  
solve the two coupled integral Eqs. (\ref{HF}). Let us derive the 
explicit expression  of  Eqs. (\ref{HF}). For simplicity, let us take 
$t_h=t_c=1$.

 The states $\psi_{\alpha}(p)$ are scattering states of energies 
$E_{\alpha}= -2 \cos k_{\alpha}$, which are inside the conduction band 
($-2 \leq E_{\alpha} \leq 2$) of the leads, and bound states below 
($E_{\alpha} <-2$) or above ($E_{\alpha} > 2$) this band. 
The contribution of the bound states to $\langle c^\dagger_0 
c^{\phantom{\dagger}}_{1} \rangle$ and $\langle c^\dagger_0 
c^{\phantom{\dagger}}_{0} \rangle$ is important, since they are 
centered inside the nano-system and decay exponentially outside.

 The wave functions of the conduction band can be written in the leads as 
\begin{equation}
\begin{aligned}
\psi_{\alpha,+}(p) = \frac{1}{\sqrt{2 \pi}}
\begin{cases}  e^{ik_{\alpha} (p-\frac{1}{2})} +  
r_{\alpha} e^{-ik_{\alpha} (p-\frac{1}{2})}  
& \text{   if    } p \leq 0 \\
 t_{\alpha} e^{ik_{\alpha} (p-\frac{1}{2})}  
&\text{    if     } p \geq 1 \end{cases} \\
\psi_{\alpha,-}(p) =  \frac{1}{\sqrt{2 \pi}}
\begin{cases} e^{- ik_{\alpha} (p-\frac{1}{2})} +  
r_{\alpha} e^{ik_{\alpha} (p-\frac{1}{2})}
&\text{    if     } p \geq 1 \\ 
    t_{\alpha} e^{- ik_{\alpha} 
(p-\frac{1}{2})}   &\text{    if     } p \leq 0 
\end{cases}
\end{aligned}
\label{wf-conduction}
\end{equation}
where 
\begin{equation}
\begin{aligned}
r_{\alpha}=& \frac {e^{ik_{\alpha}}(-1+v^2-V^2-2V\cos k_{\alpha})}
{1+2V e^{i k_{\alpha}}+(V^2-v^2) e^{2ik_{\alpha}}}          \\
t_{\alpha}=& \frac {v(e^{2ik_{\alpha}}-1)}
{-1-2Ve^{ik_{\alpha}}+(v^2-V^2) e^{2ik_{\alpha}}}.           
\end{aligned}
\label{ref-trans}
\end{equation}

 There are 4 possible bound states centered on the nano-system. 
Their wave functions take the general form: 
\begin{equation}
\psi_{bs}^{\alpha,\beta} (p)= A_{\alpha,\beta} (-1)^{p\alpha} 
\mathrm{sign} (p\beta-\frac{\beta}{2}) e^{-K_{\alpha,\beta}|p-\frac{1}{2}|} 
\label{bs}.
\end{equation} 
Only two bound states of energies $E_{\alpha,\beta}=-2 
\cosh(K_{\alpha,\beta})$ can exist below the conduction band. 
The first ($\alpha,\beta=0,1$) exists if $-(v+V)>1$ with 
$K_{0,1}=\ln(-(v+V))$. The second  ($\alpha,\beta=0,0$) exists if $v-V>1$ 
with $K_{0,0}=\ln(v-V)$. 

 From Eqs. (\ref{wf-conduction}) and (\ref{bs}) the expectation values of 
$\langle c^\dagger_0 c^{\phantom{\dagger}}_{1} \rangle$ and $\langle 
c^\dagger_0 c^{\phantom{\dagger}}_{0} \rangle$ can be explicitly calculated. 
The contributions of the conduction band read 
\begin{equation}
\begin{aligned}
\left \langle c^\dagger_0 c^{\phantom{\dagger}}_{1} \right \rangle _{cb} = & 
\sum_{q=\pm} \int_{0}^{k_F} \psi_{\alpha,q}(0)^* \psi_{\alpha,q}(1) 
dk_{\alpha} \\
= & \frac{F_{-}+2v(k_FV+\Delta \sin k_F)}{2\pi \Delta^2} \\ 
\left \langle c^\dagger_0 c^{\phantom{\dagger}}_{0} \right \rangle _{cb} = & 
\sum_{q=\pm} \int_{0}^{k_F} |\psi_{\alpha,q}(0)|^2 dk_{\alpha} \\
=&\frac{F_{+}+k_F(v^2+V^2+\Delta^2)+2V \Delta \sin k_F} {2\pi \Delta^2} 
 \end{aligned}
\label{cont-cb}
\end{equation}
respectively, where we have introduced different auxiliary functions: 
\begin{equation}
\begin{aligned}
\Delta = & v^2-V^2 \\
f_0(\pm) = & \arctan \left( \frac{v \pm (V-1)}
{v \pm (V+1)} \tan \frac{k_F}{2} \right), \\
f_{\pm}=& f_0 (\pm) \left(\Delta^2 - (v \mp V)^2 \right), \\ 
F_{\pm}=& f_{+} \pm f_{-}. 
\end{aligned}
\end{equation}
The contribution of the bound states reads: 
\begin{equation}
\begin{aligned}
\left \langle c^\dagger_0 c^{\phantom{\dagger}}_{1} \right \rangle_{bs} = & 
\left( \frac{1}{2} - \frac{1}{2(v-V)^2} \right) \Theta(v-V-1) \\ &
+\left( - \frac{1}{2} + \frac{1}{2(v+V)^2} \right) \Theta(-v-V-1) \\
\left \langle c^\dagger_0 c^{\phantom{\dagger}}_{0} \right \rangle_{bs} = & 
\left( \frac{1}{2} - \frac{1}{2(v-V)^2}\right) \Theta(v-V-1) \\ &
+\left( \frac{1}{2} - \frac{1}{2(v+V)^2} \right) \Theta(-v-V-1),
\end{aligned}
\label{cont-bs}
\end{equation}
where $\Theta(x)$ is the Heaviside step-function.

Using Eqs. (\ref{cont-cb}) and (\ref{cont-bs}) and  
\begin{equation}
\begin{aligned}
\left \langle c^\dagger_0 c^{\phantom{\dagger}}_{1} \right \rangle =& 
\left \langle c^\dagger_0 c^{\phantom{\dagger}}_{1} \right \rangle_{cb} + 
\left \langle c^\dagger_0 c^{\phantom{\dagger}}_{1} \right \rangle_{bs} \\
\left \langle c^\dagger_0 c^{\phantom{\dagger}}_{0} \right \rangle =& 
\left \langle c^\dagger_0 c^{\phantom{\dagger}}_{0} \right \rangle_{cb} + 
\left \langle c^\dagger_0 c^{\phantom{\dagger}}_{0} \right \rangle_{bs},
\end{aligned}
\label{explicit-HF}
\end{equation}
one gets an explicit form of the two integral Eqs. (\ref{HF}), which can 
be numerically solved for obtaining the self-consistent values of $v$ and 
$V$.

\subsection{Numerical method for having the HF parameters}
\label{numerical-method}
\begin{figure}
\centerline{\includegraphics[width=\linewidth]{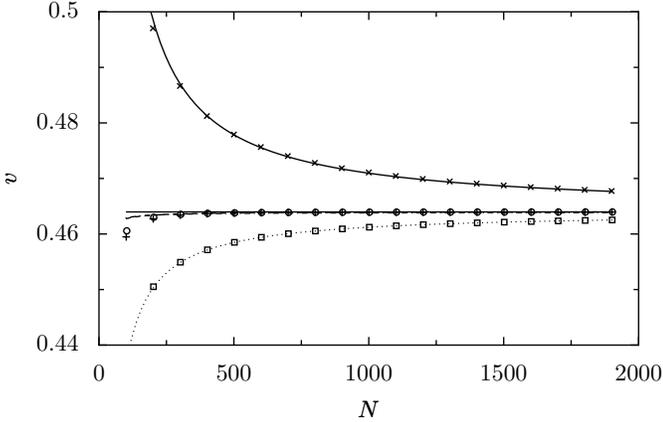}}
\caption{Effective hopping term $v$ of a nano-system with two finite 
1d leads of respective lengths $N_L$ and $N_R$ ($N_L \approx N_R$) as 
a function of the total length $N=N_L+N_R+2$, for $k_F=\pi/2$, $U=2$, 
$t_h=t_c=1$ and $t_d=0.1$.}
\label{fig5} 
\end{figure}

  If one includes other scatterers in the leads, to calculate the 
analytical form of the HF equations becomes tedious. It is faster 
to obtain  $v$ and $V$ by an alternative numerical method, based 
on the numerical diagonalization of a one body system of size $N$, 
composed of $H_S^{HF}$ coupled to two finite leads of size $N_L$ and 
$N_R$, with $N_L \approx N_R$ and $N_R+N_L+2=N$. There are 4 possible 
cases: $N_L=N_R$ or $N_L=N_R+1$ for $N_L$ even or odd. Taking consecutive 
sizes $N$, $N+1$, $N+2$ and $N+3$, this gives the 4 different curves 
shown in Fig. \ref{fig5}, which converge towards the same asymptotic value 
$v$. This asymptotic value corresponds to the value obtained from the two 
coupled integral Eqs. (\ref{HF}) with the explicit form given by 
Eqs. (\ref{explicit-HF}). 

\section{Two limits for the Hartree-Fock approximation} 
\label{section2}

\subsection{Tractable limit ($t_d>t_h$, $|t_s|^2$ independent of 
external scatterers)}

 In the limit where $t_d$ is large, such that $V_G-t_d \ll E_F$ and 
$V_G+t_d+U \gg E_F$, there is a large interval of values of $V_G$ and 
$E_F$ where the HF parameters read  
\begin{equation}
\begin{aligned}
v =& t_d+\frac{U}{2} \\ 
V= & V_G+\frac{U}{2}.   
\label{HF-trivial}
\end{aligned}
\end{equation}
For showing this, let us consider the case without interaction ($U=0$). 
The one body Hamiltonian 
$H_0=H_{s}(U=0) + \sum_{J=L,R} (H_{lead}^{J}+H_{coupling}^{J})$ 
gives rise to a $N \times N$ Hamiltonian matrix  
\begin{equation}
{\cal H}_0= \left( \begin{array} {ccc}
{\cal H}_{lead}^{L} & {\cal H}_L & 0 \\
{\cal H}_L^T & {\cal H}_{4} & {\cal H}_R \\
0 & {\cal H}_{R}^{T} & {\cal H}_{lead}^{R}
\end{array}
\right) 
\end{equation}
in the site basis, where the $4 \times 4$ matrix  
\begin{equation}
{\cal H}_{4}= \left( \begin{array} {cccc}
 0 & -t_c & 0 & 0 \\
-t_c &  V_G & -t_d & 0 \\
0 & -t_d & V_G & -t_c  \\
0 & 0 &  -t_c & 0
\end{array}
\right) 
\end{equation}
describes the nano-system with its coupling to the leads. Assuming that 
the two leads have an equal length $N_L=N_R=L$, ${\cal H}_{lead}^{L}$ 
(${\cal H}_{lead}^{R}$) are the $(L-1) \times (L-1)$ matrices describing the 
left (right) lead of size $L$ (minus its last (first) site). ${\cal H}_L$ 
(${\cal H}_R$) are $(L-1) \times 4$ ($4 \times (L-1)$) matrices with a 
single non zero matrix element $-t_h$ describing the hopping between the 
lead and its last (first) site. 

Let us introduce a $N \times N$ orthogonal transformation ${\cal O}$ which 
contains a $4\times4$ matrix 
\begin{equation}
{\cal O}_{4}= \left( \begin{array} {cccc}
 1    & 0 & 0 & 0 \\
0 &  \frac{1}{\sqrt 2} &  \frac{1}{\sqrt 2}  & 0 \\
0 &   \frac{1}{\sqrt 2} & - \frac{1}{\sqrt 2} & 0 \\
0 & 0 &  0 & 1
\end{array}
\right)
\end{equation}
acting upon ${\cal H}_{4}$, such that 
\begin{equation}
{\cal O}_{4}^{T} {\cal H}_{4}{\cal O}_{4} 
= \left( \begin{array} {cccc}
0    & -\frac{t_c}{\sqrt 2} & -\frac{t_c}{\sqrt 2} & 0 \\
-\frac{t_c}{\sqrt 2} &  V_S^0 & 0    & -\frac{t_c}{\sqrt 2}\\
-\frac{t_c}{\sqrt 2} &  0 & V_A^0 & +\frac{t_c}{\sqrt 2} \\
0 & -\frac{t_c}{\sqrt 2} & +\frac{t_c}{\sqrt 2} & 0
\end{array}
\right),  
\label{H4}
\end{equation}
where $V_A^0=V_G+t_d$ and $V_S^0=V_G-t_d$. ${\cal O}$ leaves ${\cal H}$ 
unchanged otherwise. Let us introduce the operators $d^{\phantom{\dagger}}_S=
(c^{\phantom{\dagger}}_0+c^{\phantom{\dagger}}_1)/\sqrt2$ and 
$d^{\phantom{\dagger}}_A=
(c^{\phantom{\dagger}}_0-c^{\phantom{\dagger}}_{1})/\sqrt2$, 
corresponding respectively to the symmetric (antisymmetric) 
combination of the nano-system orbitals. $n_S=d^{\dagger}_S 
d^{\phantom{\dagger}}_S$ and $n_A=d^{\dagger}_A d^{\phantom{\dagger}}_A$. 
Since $n_1n_0=n_An_S$, the HF equations (\ref{HF}) become in the transformed 
basis
\begin{equation}
\begin{aligned}
V_A = & V_A^0 + U \left \langle d^\dagger_S d^{\phantom{\dagger}}_{S} 
\right \rangle \\
V_S = & V_S^0 + U \left \langle d^\dagger_A d^{\phantom{\dagger}}_{A} 
\right \rangle \\
v_{AS} = & U \left \langle d^\dagger_A d^{\phantom{\dagger}}_{S} 
\right \rangle,   
\label{HF-trivial2}
\end{aligned}
\end{equation}
where $v_{AS}=0$, since 
\begin{equation}
\left \langle d^\dagger_A d^{\phantom{\dagger}}_{S} \right 
\rangle =\frac{1}{2} \left ( \left \langle n_0  \right \rangle - 
\left \langle n_1 \right \rangle+ \left \langle 
c^{\dagger}_0 c^{\phantom{\dagger}}_1 \right \rangle - 
\left \langle 
c^{\dagger}_1 c^{\phantom{\dagger}}_0 \right \rangle \right ), 
\end{equation}
is equal to zero if the system is invariant under the inversion 
$0 \leftrightarrow 1$. 

\begin{figure}
\centerline{\includegraphics[width=\linewidth]{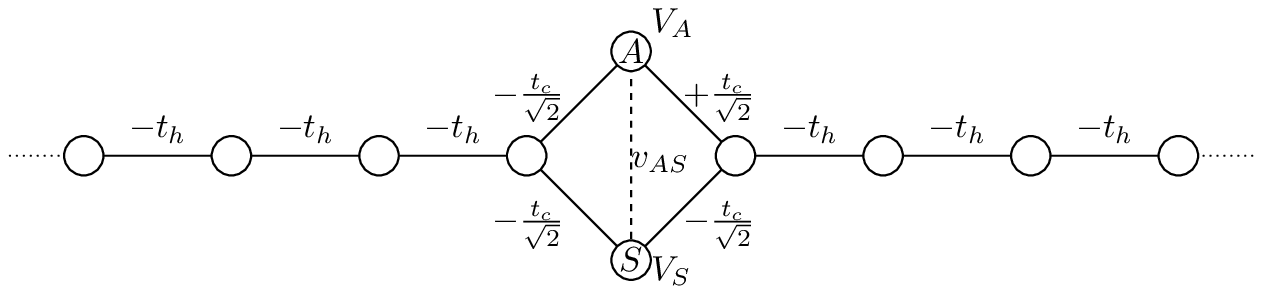}}
\caption
{
Equivalent set-up obtained by the orthogonal transformation $\cal O$ 
from the original set-up drawn in Fig. \ref{fig4}. The nano-system 
is now made of two sites in parallel connected to the 2 leads by 
modified hopping terms $\pm t_c/{\sqrt 2}$. The site corresponding 
to the symmetric (anti-symmetric) orbital has an energy $V_S$ ($V_A$) 
which is given by Eqs. (\ref{HF-parameter-trivial}). The hopping term 
$v_{AS}$ due to exchange is zero when there is reflection symmetry.
}
\label{fig6} 
\end{figure} 

  The equivalent set-up obtained by the orthogonal transformation 
$\cal O$ from the original set-up is sketched in Fig. \ref{fig6}. There 
are three simple limiting cases: two correspond to the limit where either 
$V_A, V_S \ll E_F$ or $V_A,V_S \gg E_F$, such that the two sites of the 
nano-system are either totally  filled or totally empty. This yields an 
effective transmission $|t_s|^2 \approx 0$ at $E_F$. The third case 
corresponds to a site $A$ (anti-symmetric orbital) with an occupation number 
$\langle n_A \rangle \approx 0$ ($V_A \gg E_F$) and a site $S$ (symmetric 
orbital) with $\langle n_S \rangle \approx 1$ ($V_S \ll E_F$). The larger 
is $t_d$, the larger is the range of values of $V_G$ corresponding to this 
limit, for a given Fermi energy $E_F$. In that case, Eqs. (\ref{HF-trivial2}) 
give
\begin{equation}
\begin{aligned}
V_A = & V_G+t_d+U \\
V_S = & V_G-t_d.
\label{HF-parameter-trivial}
\end{aligned}
\end{equation}
Putting in the $4 \times 4$ matrix given by Eq. (\ref{H4}) those HF values  
$V_A$ and $V_S$ instead of the bare values $V_A^0$ and $V_S^0$ defines 
${\cal H}_{4}^{HF}$. Calculating ${\cal O}_{4} {\cal H}_{4}^{HF}
{\cal O}_{4}^{T}$, one finds for the HF parameters $V$ and $v$ the values 
given by Eqs. (\ref{HF-trivial}).
 
\begin{figure}
\centerline{\includegraphics[width=\linewidth]{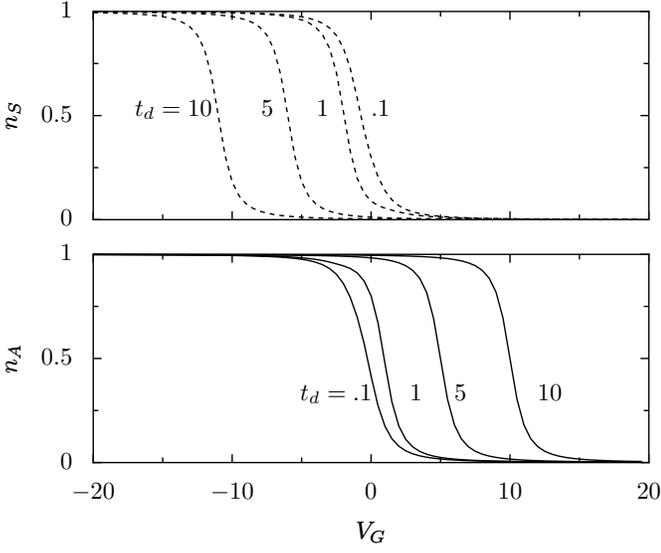}}
\caption
{
Occupation numbers  $\langle n_S \rangle$ (solid lines) and  
$\langle n_A \rangle$ (dashed lines) as a function of $V_G$ for 
$k_F=\pi/8$ and different values of $t_d$ given in the figure. 
$U=1$, $t_c=t_h=1$. 
}
\label{fig7} 
\end{figure} 
\begin{figure}
\centerline{\includegraphics[width=\linewidth]{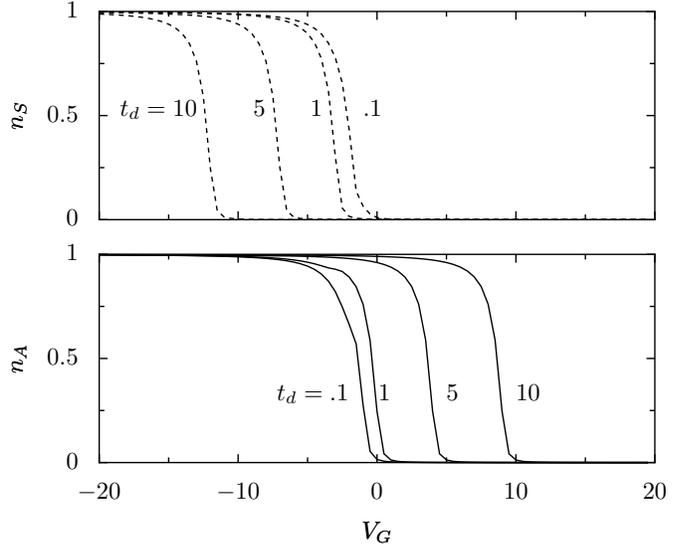}}
\caption
{Occupation numbers  $\langle n_S \rangle$ (solid lines) and  
$\langle n_A \rangle$ (dashed lines) as a function of $V_G$ for 
$k_F=\pi/2$ and different values of $t_d$ given in the figure. 
$U=1$, $t_c=t_h=1$.
}
\label{fig8} 
\end{figure} 

 Using the analytical form of HF equations given in subsection \ref{AFHFE}, 
we have calculated the two occupation numbers $\langle n_S \rangle$ and 
$\langle n_A \rangle$ as a function of $V_G$ for different values of 
$E_F=-2 \cos k_F$ and $t_d$. The results are shown assuming a nano-system 
well coupled to the leads ($t_c=t_h=1$), for $k_F=\pi/8$ (Fig. \ref{fig7}) 
and for $k_F=\pi/2$ (Fig. \ref{fig8}). One can see that for $t_d \gg 1$, 
there are large intervals of values of $V_G$ for which 
$\langle n_S \rangle \approx 1$ and $\langle n_A \rangle \approx 0$. 
In that case, $v$ and $V$ are given by Eqs. (\ref{HF-trivial}) and it is 
very easy to obtain the nano-system transmission $|t_s|^2$ at 
the Fermi energy $E_F$. For renormalized hopping term $v$ and gate 
potential $V$, the effective transmission reads  
\begin{equation}
|t_s|^2 \approx  \frac{v}{x} \left 
( \frac{\Gamma^2}{(v-x)^2 + \Gamma^2} - 
\frac{\Gamma^2}{(v+x)^2 + \Gamma^2} \right), 
\label{transmissionHF}
\end{equation}
where 
\begin{equation}
\begin{aligned}
\Gamma = & t_c^2 \sin k_F \\ 
x = & V-(t_c^2-2) \cos k_F. 
\end{aligned}
\end{equation} 
If $v$ and $V$ are given by Eqs. (\ref{HF-trivial}), one finds:  
\begin{equation}
|t_s|^2 \approx  \Delta \left ( \frac{\Gamma^2}{(V_G-V_1)^2 - \Gamma^2} - 
\frac{\Gamma^2}{(V_G-V_2)^2 - \Gamma^2} \right), 
\label{transmissionHF-trivial}
\end{equation}
where
\begin{equation}
\begin{aligned}
\Delta = & \frac{2t_d+U}{2V_G+U-2 (t_c^2-2) \cos k_F} \\
V_1 = & t_d+(t_c^2-2) \cos k_F \\ 
V_2 = & -t_d +(t_c^2-2) \cos k_F -U.
\end{aligned}
\end{equation} 

When one varies $V_G$, Eq. (\ref{transmissionHF-trivial}) gives for 
the transmission $|t_s|^2$ two transmission peaks located at $V_G = V_1$ 
and $V_2$, and spaced by a large interval $2t_d+U$ when $t_d$ is large. 

When $t_c \ll 1$, $(t_c^2-2) \cos k_F \approx E_F$ and the nano-system is 
very weakly coupled to the leads, with two levels of energy $V_G \pm t_d$ 
when $U=0$. There are two sharp transmission peaks of width 
$\Gamma = t_c^2 \sin k_F \ll 1$, the first when $E_F \approx V_G-t_d$ 
($V_G=V_1$), the second when $E_F\approx V_G+t_d+U$ ($V_G=V_1$). 
Since one needs an energy $E_F$ for putting an electron outside the 
nano-system, and an energy $V_G-t_d$ to put an electron inside the 
empty nano-system, or $V_G+t_d+U$ inside the nano-system occupied by another 
electron, one recovers the usual Coulomb Blockade, where the nano-system 
has a transmission peak when it is indifferent for an electron to be 
inside or outside the nano-system. 

When $t_c \rightarrow 1$, the nano-system becomes  strongly coupled to the 
leads, the peak width $\Gamma$ is broader and the two values of $V_G$ for 
which the transmission is large are shifted by an amount equal to $E_F/2$.

\begin{figure}
\centerline{\includegraphics[width=\linewidth]{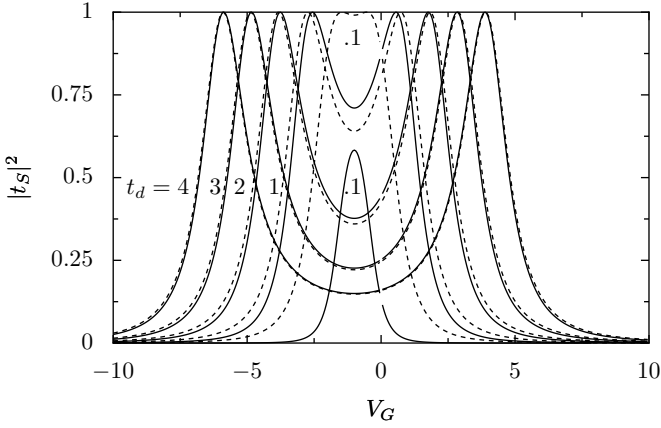}}
\caption
{
Effective nano-system transmission $|t_s|^2$ as a function of 
$V_G$ for $k_F=\pi/2$ and different values of $t_d$ given in the 
figure. $U=2$, $t_c=t_h=1$. The solid lines give $|t_s|^2$ calculated 
using Eq. (\ref{transmissionHF}) with the HF parameters $v$ and $V$ 
calculated exactly. The dashed lines give $|t_s|^2$ calculated 
using Eq. (\ref{transmissionHF-trivial}) ($v$ and $V$ given by 
Eqs. (\ref{HF-trivial})).
}
\label{fig9} 
\end{figure}

\begin{figure}
\centerline{\includegraphics[width=\linewidth]{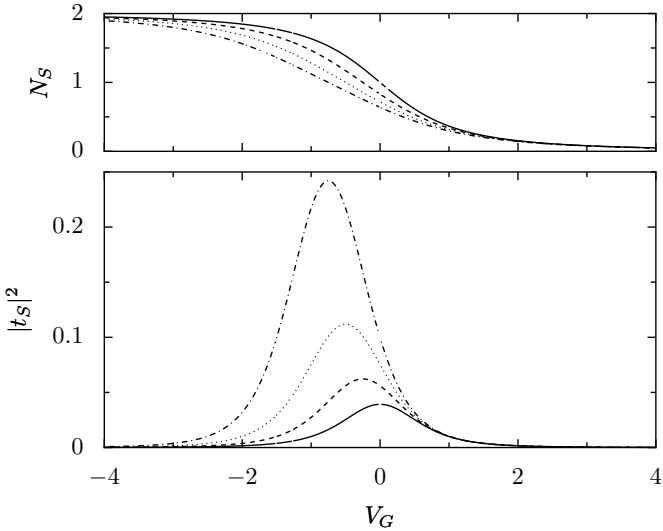}}
\caption
{
$t_d=0.1$, $k_F=\pi/2$, $t_c=t_h=1$.
Nano-system occupation number $N_S=\langle n_A \rangle + 
\langle n_S \rangle$ (up) and corresponding effective 
transmission $|t_s|^2$ (down) as a function of $V_G$ for 
$U=0$ (solid), $0.5$ (dashed) $1$ (dotted) $1.5$ (dashed-dotted).
}
\label{fig10} 
\end{figure}

This double peak structure is shown in Fig. \ref{fig9} when $t_d \gg 
t_h$ and $t_c=t_h=1$. It agrees with the curve given by 
Eq. (\ref{transmissionHF-trivial}). In contrast, this approximation 
totally fails to describe the single peak structure occurring when $t_d=0.1$, 
as shown in Fig. \ref{fig9} and Fig. \ref{fig10}. 


When $t_d$ is large, the symmetric site $S$ of potential $V_S=V_G-t_d$ 
is far from the anti-symmetric site of potential $V_A=V_G+t_d+U$. If 
the nano-system is empty ($|t_s|^2\approx 0$) and if one varies $E_F$ 
for a given value of $V_G$, or $V_G$ for a given value of $E_F$, one 
first fills the symmetric state, then the anti-symmetric one. This gives 
two transmission peaks.  When the two potentials $V_A$ and $V_S$ are far 
from $E_F$, $\langle n_A \rangle$ and  $\langle n_S \rangle$ are either 
$0$ or $1$, and only huge external Friedel oscillations could enter 
inside the nano-system and vary  $|t_s|^2$. In that case, the nano-system 
occupation number $N_S=\langle n_A \rangle + \langle n_S \rangle $ is 
locked to values $0$, $1$ or $2$, which cannot be changed by external 
Friedel oscillations. This makes the sensitivity of $|t_s|^2$ to external 
scatterers very negligible in that case. If one of the two 
renormalized potentials $V_A$ and $V_S$ is near $E_F$, only the component 
of external Friedel oscillations with the right symmetry can go through 
the equivalent site $A$ or $S$ of the same symmetry. Even in that case, 
the change of $|t_s|^2$ by an external scatterer cannot be very large. 
The limit where the solution of HF equations is straightforward is also 
the limit where the nano-system transmission is almost independent of 
external scatterers.

\subsection{Non local Limit ($t_d < t_h$, $|t_s|^2$ dependent of 
external scatterers)}

 When $t_d$ is small, the symmetric site $S$ of potential $V_G-t_d$ and 
the anti-symmetric site $A$ of potential $V_G+t_d+U$ can be put together 
near $E_F$ by a suitable strength of $V_G$. In that case, the two 
transmission peaks merge into a single one, as shown in Figs. \ref{fig9} 
and \ref{fig10} for $t_d=0.1$. Looking in Figs. \ref{fig7} and 
\ref{fig8}, one can see that $\langle n_A \rangle \approx \langle n_S 
\rangle$ take intermediate values between $0$ and $1$ around this single 
transmission peak, the potentials $V_A$ and $V_S$ being near $E_F$ for 
the same values of $V_G$. This is the interesting limit where one can 
strongly vary $\langle n_A \rangle $ and $\langle n_S \rangle$ by external 
scatterers, the induced Friedel oscillations being able to enter inside the 
nano-system. Large variations of the HF parameters can be expected in this 
limit, and hence large changes of the effective transmission $|t_s|^2$. If 
the external scatterer is made of an attached ring, the induced Friedel 
oscillations entering inside the nano-system can be changed by an AB flux 
threading the ring, and $|t_s|^2$ can exhibit large AB oscillations. 

\section{Aharonov-Bohm scatterer} 
\label{section3}

  The AB-scatterer sketched in Fig. \ref{fig1} contains an attached ring, 
such that it can induce flux dependent Friedel oscillations in the lead when 
an AB flux is varied through the ring. Its topology requires two 3-lead 
contacts (3LC).  A 3LC is made of 4 coupled sites indicated by black circles 
in Fig. \ref{fig1}, and is described by a local Hamiltonian
\begin{equation}
H_P=\sum_{p=1}^3  -t_{p} (c^\dagger_P c^{\phantom{\dagger}}_{p} 
+ h.c),
\end{equation}
$P$ denoting the central site and the sum $p$ being taken over its 
3 neighbors. The hopping terms are taken equal $t_{p}=t_{h}=1$. 
The first 3LC allows us to attach a vertical lead to the horizontal lead, 
the second one to attach the ring to this vertical lead. $L_c$ is the 
number of sites between the upper 3LC and the nano-system. Varying  
$L_c$, one can study the influence of the AB scatterer upon the effective 
transmission $|t_s|^2$ of the nano-system. $L'_c$ and $L_R$ are 
respectively the numbers of sites between the two 3LCs (length of the 
vertical lead) and of the attached ring (length of the ring without 
the three sites of the lower 3LC), as shown in Fig. \ref{fig1}. 
A $3 \times 3$ matrix $S_P(k)$ describes the scattering by a 3LC at an 
energy $E=-2 \cos k$: 
\begin{equation}
S_P(k)= \left( \begin{array} {ccc}
s_{d} & s_{o} & s_{o}  \\
s_{o} & s_{d} & s_{o}  \\
s_{o} & s_{o} & s_{d} 
\end{array}
\right)
\end{equation}
where 
\begin{equation}
\begin{aligned}
s_{d}=& \frac{- e^{ik}}{3 e^{ik}-2 \cos k} \\
s_{o}=& \frac{ 2i \sin k}{3 e^{ik}-2 \cos k}.
\end{aligned}
\end{equation}

The reflection amplitude of an incoming electron of the vertical lead 
by the ring threaded by a flux $\Phi$ reads 
\begin{equation}
r_R(\varphi)=\frac{h_k(\varphi) - \sin (kL_R)}{-h_k(\varphi)+e^{2ik} 
\sin (kL_R)},
\label{boucle}
\end{equation}
where 
\begin{equation}
h_k(\varphi)=2 e^{ik} (\cos (kL_R)- \cos \varphi) \sin k
\end{equation}
$\phi = 2\pi\Phi/\Phi_0$, $\Phi_0$ being the flux quantum. 

The reflection and transmission amplitudes of an electron moving 
in the horizontal lead by the AB-scatterer read
\begin{equation}
r_{AB}(k)=\frac{-e^{2ik}-e^{2ikL'_{c}} r_R(\varphi)}
{2 e^{2ik}-1+r_R(\varphi) e^{2ik(L'_c+1)}}
\label{r_{AB}}
\end{equation}
\begin{equation}
t_{AB}(k)=\frac{2i \sin k e^{ik} (1+e^{2ikL'_{c}}r_R(\varphi))} 
{2e^{2ik}-1 +r_R(\varphi)e^{2ik(L'_c+1)}}
\label{t_{AB}}.
\end{equation}

\section{Friedel oscillations and particle-hole symmetry} 
\label{section4}
 
 If one puts a symmetric nano-system in series with an AB-scatterer,
the inversion symmetry is broken, and the potentials $V_0 \neq 
V_1$.  In that case, one has to calculate the values $v$, $V_0$ and 
$V_1$ of the HF parameters satisfying the three coupled HF equations
\begin{equation}
\begin{aligned}
v=& t_d+U\left \langle c^\dagger_0 c^{\phantom{\dagger}}_{1} (v,V_0,V_1)
\right \rangle \\
V_0=& V_G+U\left\langle c^\dagger_1 c^{\phantom{\dagger}}_{1} (v,V_0,V_1) 
\right \rangle \\
V_1=& V_G+U\left\langle c^\dagger_0 c^{\phantom{\dagger}}_{0} (v,V_0,V_1)
\right \rangle,  
\label{hf2}
\end{aligned}
\end{equation}
instead of the two HF Eqs. (\ref{HF}) valid when $V_0=V_1$. 

 The non local effect is a consequence of the corrections to 
$\langle c^\dagger_0 c^{\phantom{\dagger}}_{1} \rangle$, $\langle 
c^\dagger_0 c^{\phantom{\dagger}}_{0} \rangle$ and $\langle c^\dagger_1 
c^{\phantom{\dagger}}_{1} \rangle$ which are induced inside the nano-system 
by the AB scatterer. In the general case, the AB scatterer and the 
nano-system induce at a site $p$ Friedel oscillations of the 
density $\langle c^\dagger_p c^{\phantom{\dagger}}_{p} \rangle$ and 
similar oscillations of the correlation function $\langle c^\dagger_p 
c^{\phantom{\dagger}}_{p+1} \rangle$. Let us illustrate the effect of 
each scatterer inside the attached leads when there is particle-hole 
symmetry. In this particular case, the density stays uniform, $\langle 
c^\dagger_p c^{\phantom{\dagger}}_{p} \rangle=1/2$ everywhere and there 
are no Friedel oscillations of the density. But the effect of the AB 
scatterer upon the nano-system transmission $|t_s|^2$ persists, because 
of the exchange contribution, and one just needs to study $\langle 
c^\dagger_p c^{\phantom{\dagger}}_{p+1} \rangle$. Particle-hole symmetry 
occurs at half-filling ($k_F=\pi/2$) when one takes a gate potential 
$V_G=-U/2$ which exactly compensates the Hartree contributions $U/2$, 
such that $V_0=V_1=0$. 

\subsection{Interaction dependent oscillations induced by the nano-system}

\begin{figure}
\centerline{\includegraphics[width=\linewidth]{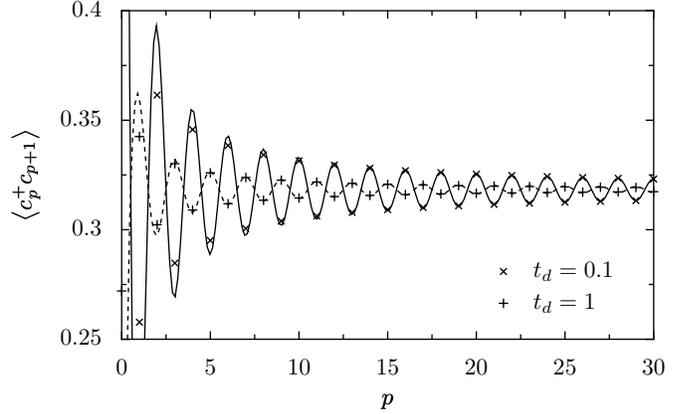}}
\caption
{
Even-odd oscillations of $\langle c^\dagger_p c_{p+1} \rangle$ towards the 
asymptotic value $1/\pi$, induced inside the leads 
by the interacting nano-system for $k_F=\pi/2$, $U=1$, $V_G=-U/2$ and 
$t_c=t=1$. The dashed and solid lines give two asymptotic fits $1/\pi+ 
b \cos(\pi p+c)/p$ with ($b=0.04151$, $c=\pi$) and ($b=0.14776$, $c=0$) 
for $t_d=1$ (+) and $t_d=0.1$ (x) respectively.
}
\label{fig11} 
\end{figure}

In a case where particle-hole symmetry yields a uniform density, 
the usual Friedel oscillations are absent, and the exact form of 
$\langle c^\dagger_p c^{\phantom{\dagger}}_{p+1} \rangle$ is given in 
Ref. \cite{AFP} for $V_G=-U/2$ and $t_d=1$. It has an asymptotic behavior 
which reads
\begin{equation}
\left \langle c^\dagger_p c^{\phantom{\dagger}}_{p+1} \right 
\rangle \approx a + b \frac{\cos (2 k_F p +c)}{p},
\label{oscillation}
\end{equation}
where the asymptotic value $a=\sin k_F/\pi$ and the phase $c=0$ at  
$k_F=\pi/2$. This gives even-odd oscillations with a $1/p$-decay towards 
the asymptotic value $1/\pi$ which are shown in Fig. \ref{fig11} for 
$t_d=1$ and $t_d=0.1$. As expected, the amplitude $b=0.14776$ is larger 
when $t_d=0.1$ than when $t_d=1$ ($b=0.04151$). The asymptotic form given by 
Eq. (\ref{oscillation}) characterizes also the Friedel oscillations of 
$\langle c^\dagger_p c^{\phantom{\dagger}}_{p} \rangle$ when particle-hole 
symmetry is broken.

\subsection{Flux dependent oscillations induced by the AB scatterer}

\begin{figure}
\centerline{\includegraphics[width=\linewidth]{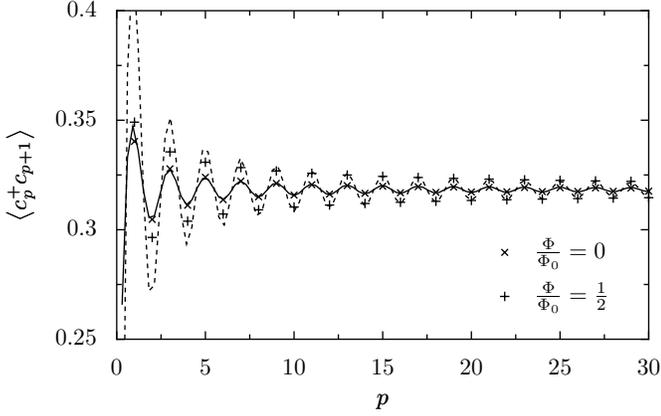}}
\caption
{
Flux dependent oscillations of $\langle c^\dagger_p c^{\protect 
\phantom{\dagger}}_{p+1} \rangle$ towards the asymptotic value $1/\pi$, 
induced by an AB scatterer with a ring of size $L_R=6$, for $\Phi=\Phi_0/2$ 
(+) or $\Phi=0$ (x) ($k_F=\pi/2$ and $L'_c=4$). The dashed 
and solid lines give two asymptotic fits $1/\pi+ b \cos(\pi p+c)/p$ with 
($b=0.09983, c=\pi$) and ($b=0.02746, c=\pi$) for $\Phi=\Phi_0/2$ and 
$\Phi=0$ respectively. 
}
\label{fig12} 
\end{figure}

\begin{figure}
\centerline{\includegraphics[width=\linewidth]{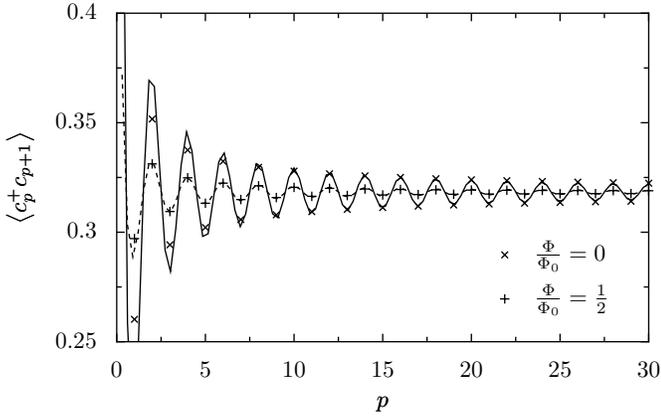}}
\caption
{Flux-dependent oscillations of $\langle c^\dagger_p c^{\protect 
\phantom{\dagger}}_{p+1} \rangle$ towards the asymptotic value $1/\pi$, 
induced by an AB scatterer with a ring 
of size $L_R=7$ for $\Phi=\Phi_0/2$ (+) and $\Phi=0$ (x) respectively , 
for $k_F=\pi/2$ and $L'_c=4$. The dashed and solid lines give two asymptotic 
fits $1/\pi+ b \cos(\pi p+c)/p$ with ($b=0.027997$, $c=0$) and ($b=0.11029$, 
$c=0$) for $\Phi=\Phi_0/2$ and $\Phi=0$ respectively. 
}
\label{fig13} 
\end{figure}

  The AB scatterer induces also flux dependent oscillations of 
$\langle c^\dagger_p c^{\phantom{\dagger}}_{p+1} \rangle$ around it, 
even though $\langle c^\dagger_p c^{\phantom{\dagger}}_{p} \rangle =1/2$ 
everywhere if $k_F=\pi/2$. These oscillations have also the asymptotic 
behavior given by Eq. (\ref{oscillation}), as shown in Fig. \ref{fig12} and 
\ref{fig13} for even and odd sizes $L_R$ of the ring ($L_R=6$ and $7$). 
At $k_F=\pi/2$, the scattering matrix elements of the AB-scatterer 
given by  Eqs (\ref{r_{AB}}) and (\ref{t_{AB}}) are independent of $\Phi$ 
when $L_R$ is even, and depend on $\Phi$ when $L_R$ is odd. However 
$\langle c^\dagger_p c^{\phantom{\dagger}}_{p+1} \rangle$ oscillates and 
varies as a function of $\Phi$ both for even and odd values of $L_R$, as 
shown in Fig. \ref{fig12} and Fig. \ref{fig13}.


\section{Role of the AB flux upon the nano-system transmission $|t_s|^2$} 
\label{section5}

\begin{figure}
\centerline{\includegraphics[width=\linewidth]{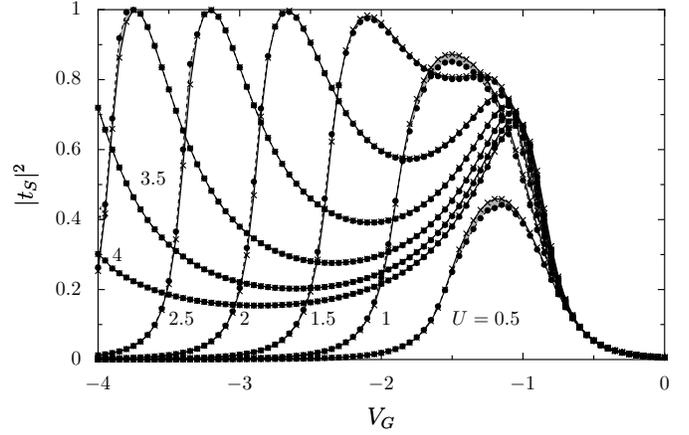}}
\caption{
Effective transmission $|t_s|^2$ as a function of the gate voltage $V_G$, 
at a filling $1/8$ (Fermi momentum $k_F=\pi/8$) and a nano-system hopping 
term $t_d=0.1$. The AB-scatterer with its attached ring 
($L'_c=4, L_R=7$) is at $L_c=2$ sites from the nano-system. The interaction 
strength $U$ is indicated in the figure. A flux $\Phi=0$ ($\bullet$) or 
$\Phi=\Phi_0/2$ (x) threads the ring. The grey areas underline the effect 
of $\Phi$ upon $|t_s|^2$.
}
\label{fig14}
\end{figure}

  When the two scatterers are put in series, the oscillations of the first 
interfere with the oscillations of the second, and the solutions of 
Eqs. (\ref{hf2}) have to be determined self-consistently. To calculate 
analytically as in subsection \ref{AFHFE} $\langle c^\dagger_0 
c^{\phantom{\dagger}}_{1} \rangle$, $\langle c^\dagger_0 
c^{\phantom{\dagger}}_{0} \rangle$ and $\langle c^\dagger_1 
c^{\phantom{\dagger}}_{1} \rangle$ becomes complicated in the presence of 
the AB scatterer. It is simpler to obtain $v$, $V_0$ and $V_1$ using the 
numerical method given in subsection \ref{numerical-method}. Once $v$, 
$V_0$ and $V_1$ are known, the effective transmission amplitude $t_s$ at an 
energy $E=-2 \cos k$ is given by 
\begin{equation}
t_{s}(k)=\frac{-2i e^{2ik} t_c^2 v \sin k } 
{F(V_0) F(V_1) - v^2},
\label{t_{s}}
\end{equation}
where $F(V)=2 \cos k + V - e^{ik} t_c^2$. 
\begin{figure}
\centerline{\includegraphics[width=\linewidth]{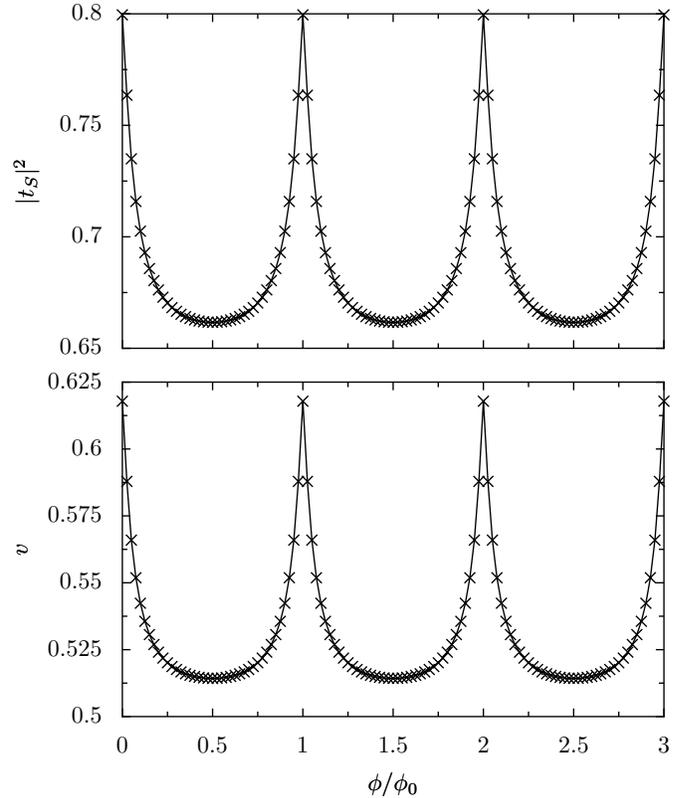}}
\caption{ Effective transmission $|t_s|^2$ (upper figure) 
and renormalized hopping $v$ (lower figure) as a function of 
$\Phi/\Phi_0$, for $k_F=\pi/2$, $U=2$ and $V_G=-1$. Same values 
as in Fig. \ref{fig2} ($L_c=2$, $L'_c=4$ and $L_R=7$ and 
$t_d=0.1$). Particle-hole symmetry ($k_F=\pi/2$, 
$V_G=-U/2$) gives $V_0=V_1=0$.
}
\label{fig15}
\end{figure}
\begin{figure}
\centerline{\includegraphics[width=\linewidth]{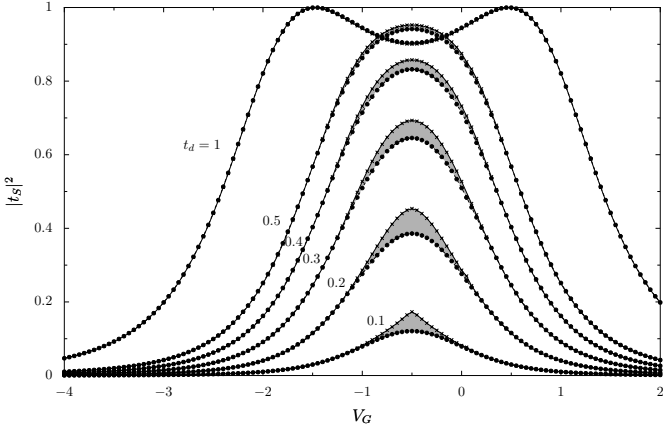}}
\caption{
Effective transmission $|t_s|^2$ as a function of $V_G$ for different 
values of the nano-system hopping term $t_d$ indicated in the figure. 
$k_F=\pi/2$, $L_c=2$, $L'_c=4$, $L_R=7$, $U=1$. A flux $\Phi=0$ ($\bullet$) 
or $\Phi=\Phi_0/2$ (x) threads the ring. The grey areas underline the effect 
of $\Phi$ upon $|t_s|^2$.
}
\label{fig16}
\end{figure}

 For having large effects of the AB flux $\Phi$ upon $|t_s|^2$, we have 
taken a small value $t_d=0.1$ for the nano-system hopping term. The results 
are shown as a function of $V_G$ in Fig. \ref{fig14} for $k_F=\pi/8$. The 
ring is attached $L_c=2$ sites away from the nano-system. The effect of 
$\Phi$ upon $|t_s|^2$ is indicated in Fig. \ref{fig14} as in 
Fig. \ref{fig2}, by grey areas between the curves $|t_s(V_G)|^2$ obtained 
with $\Phi=0$ and $\Phi=\Phi_0/2$. The effect can be seen, but remains 
small for $k_F=\pi/8$. The period $\lambda_F/2=8$ of the Friedel 
oscillations being larger than the nano-system size, it is likely that a 
stronger effect occurs if this period is reduced and becomes of the order 
of the nano-system size, when $\lambda_F/2=2$. This is confirmed in the 
Fig. \ref{fig2} which we have put in the introduction. Those large effects 
are the result of the $\Phi$-dependence of $v$, $V_0$ and $V_1$. The 
effect being particularly large in Fig. \ref{fig2} when $V_G=-1$ and 
$U=2$, we show in Fig. \ref{fig15} the corresponding AB oscillations 
characterizing $|t_s|^2$ and $v$ when $\Phi$ varies through the ring. 
As shown in Fig. \ref{fig2}, $|t_s|^2$ takes its largest value when 
$V_G=-U/2$, as far as $U$ is not too large and does not split the 
transmission peak. At $k_F=\pi/2$, this value of $V_G$ yields particle-hole 
symmetry. Therefore, the transmission is maximum when $V_G$ compensates the 
Hartree terms, such that $V_0=V_1=0$ without the AB-scatterer, the only 
source of scattering being due to the hopping term $v  \neq t_h$. One can 
also see in  Fig. \ref{fig2} that the largest dependence of $|t_s|^2$ upon 
$\Phi$ occurs for $V_G=-U/2$ at $k_F=\pi/2$.    

  The role of $t_d$ upon the strength of the non local effect is 
illustrated in Fig. \ref{fig16} for $k_F=\pi/2$ and $U=1$. The dependence 
of $\Phi$  upon $|t_s|^2$ cannot be seen at the used scale when $t_d=1$. 
This is also a value of $t_d$ where Eq. (\ref{transmissionHF-trivial}) 
gives a good approximation of $|t_s|^2$ (see Fig. \ref{fig9}). When 
$t_d$ decreases, the grey areas underlining the role of $\Phi$ upon 
$|t_s|^2$ increase around $V_G=-U/2$, where there is particle-hole 
symmetry. Of course, $|t_s|^2 \rightarrow 0$ as $t_d \rightarrow 0$.

\section{Quantum conductance $g_T$}
\label{section6}

\begin{figure}
\centerline{\includegraphics[width=\linewidth]{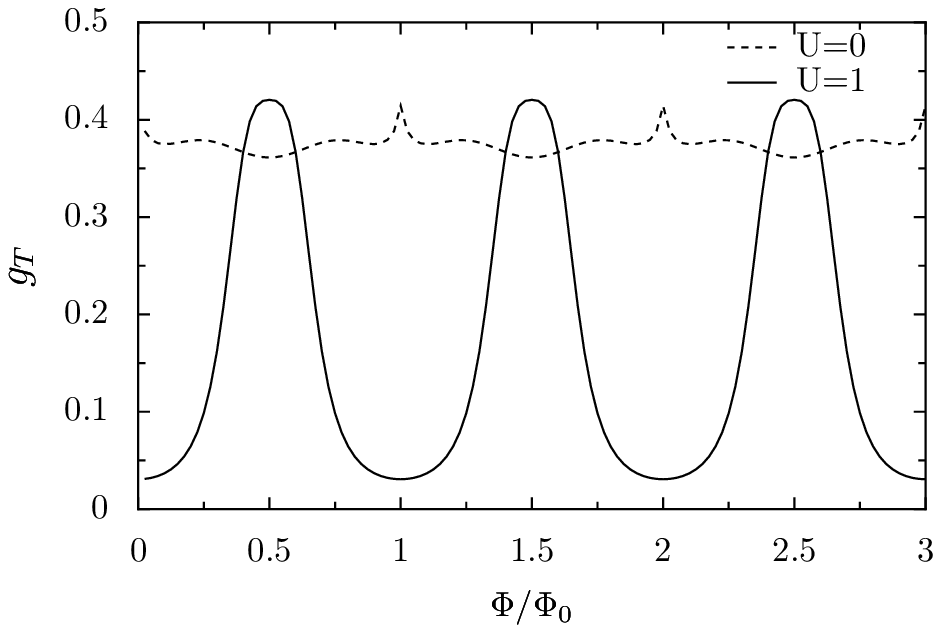}}
\caption{
Quantum conductance $g_T$ of the nano-system and the AB-scatterer 
in series as a function of $\Phi/\Phi_0$, when $U=0$ (dotted line) and 
$U=1$ (solid line). $L_c=4$, $L_R=7$, $L'_c = 6$, $V_G=-0.5$ and 
$k_F=\pi/2$. The AB-oscillations occurring without interaction 
($\sin (k_FL_R) \neq 0$) are strongly increased when $U=-2 V_G$. 
}
\label{fig17}
\end{figure}

\begin{figure}
\centerline{\includegraphics[width=\linewidth]{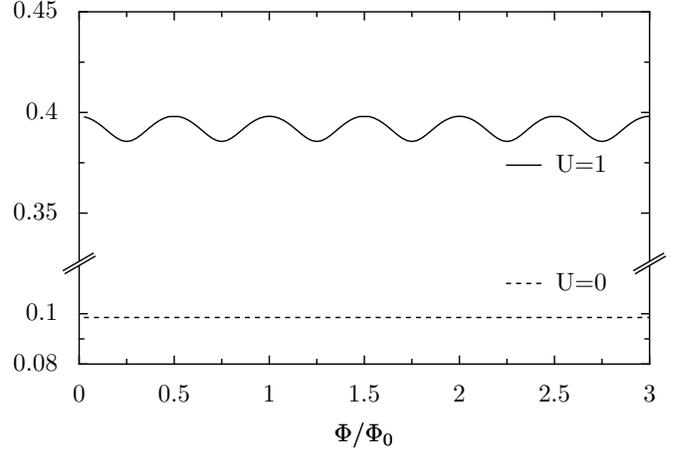}}
\caption{
$g_T$ as a function of $\Phi/\Phi_0$, when $U=0$ (dotted line) and $U=1$ 
(solid line). $L_c=4$, $L_R=6$, $L'_c = 5$, $V_G=-0.5$ and $k_F=\pi/2$. 
Without interaction, there are no AB-oscillations ($\sin (k_FL_R)=0$). 
The interaction inside the nano-system increases $g_T$ when $U=-2 V_G$ 
and yields AB-oscillations. 
}
\label{fig18}
\end{figure}

  In the two probe geometry described by Fig. \ref{fig1}, the quantum 
conductance $g_T$ of the nano-system and the AB-scatterer in series is 
given by Landauer formula which reads $g_T=|t_T(E_F)|^2$ (in units of 
$e^2/h$) in the limit where the temperature $T \rightarrow 0$. 
Using the HF approximation, the nano-system becomes an effective one body 
scatterer when $T \rightarrow 0$ and the total transmission amplitude 
$t_T(E_F)$ is given by the combination law valid for one body scatterers: 
\begin{equation}
t_T(E_F)=t_s(E_F) \frac{e^{ik_{F}L_c}}{1- r'_s(E_F) r_{AB}(E_F) 
e^{2ik_{F}L_c}} t_{AB}(E_F) .
\end{equation}
$r'_s(E_F)$ ($r_{AB}(E_F)$) is the reflection amplitude of the nano-system 
(of the AB-detector) at $E_F$. Because $r_{AB}(E_F)$ and $t_{AB}(E_F)$ 
depend in general on $\Phi$, $g_T(E_F)$ exhibits AB-oscillations even without 
interaction or if $L_c$ is very large, limits where $t_s(E_F)$ and 
$r'_s(E_F)$ are independent of $\Phi$. However, when the electrons interact 
inside the nano-system and if $L_c$ is not too large, $t_s(E_F)$ and 
$r'_s(E_F)$ exhibit also AB-oscillations which can be important around 
certain values of $V_G$ and which can strongly modify the AB-oscillations 
of the total conductance $g_T$. This is shown in Fig. \ref{fig17} for a case 
where the AB-oscillations of $g_T$ are weak without interaction, and become 
important when the electrons interact inside the nano-system. However, 
since our model depends on many parameters, it is difficult to draw 
a simple conclusion. There are also values of those parameters for which 
the AB-oscillations are large without interaction, the interaction reducing 
$g_T$ and its oscillations. 

 In our model, there are also special cases where $\sin (k_{F} L_R)=0$, 
such that the ring is perfectly reflecting and the AB-scatterer becomes 
independent of $\Phi$ at $E_F$. We show such a case in Fig. \ref{fig18} 
where $k_F=\pi/2$ and $L_R=6$, for which the interaction increases the value 
of $g_T$ (the Hartree terms compensating the value of $V_G$ when $U=-2V_G$) 
and yields AB-oscillations which are a pure many body effect. This is 
because the AB-scatterer is independent of $\Phi$ only at $E_F$, but not 
below  $E_F$. Therefore the HF parameters, and hence $t_s(E_F)$ and 
$r'_s(E_F)$, have AB-oscillations which are responsible for the 
AB-oscillations of $g_T$, while $t_{AB}(E_F)$ and $r_{AB}(E_F)$ are 
independent of $\Phi$ for $k_{F} L_R=n \pi$. 

\section{Conclusion}
\label{section7}

  In summary, we have found an effect of electron-electron interactions 
upon quantum transport, using the scattering approach to transport and the 
Hartree-Fock approximation. The study was restricted to the 1d limit with 
a temperature $T \rightarrow 0$ and spin polarized electrons. We have 
shown that the HF description of a double site nano-system becomes trivial 
if $t_d>t_h$, while the electron density inside the nano-system 
can become very sensitive to external scatterers if $t_d < t_h$. 
This is also if $t_d < t_h$ that it becomes possible to strongly 
vary the effective nano-system transmission by external scatterers.
The external scatterer which we have considered contains a ring, and can give 
rise to flux dependent Friedel oscillations if the flux through the 
ring is varied. We have shown that those long range Friedel oscillations 
can induce AB oscillations of the effective transmission, though the ring 
is attached at a distance $L_c$ from the nano-system. As explained in 
Ref. \cite{AFP}, this non local effect vanishes if the distance between the 
nano-system and the external scatterer exceeds the thermal length $L_T$ 
(length upon which an electron propagates at the Fermi velocity during a 
time $\hbar/kT$). 

  It will be of course very interesting to observe this many-body effect 
in a transport measurement. As it is well known, the strength of the 
interaction becomes more important when the electron density is reduced, 
the Coulomb to kinetic energy ratio (factor $r_s$) becoming large. A 
possibility is to take for the interacting nano-system a quantum dot where 
the electron density can be reduced by an electrostatic gate, creating a 
small region of large factor $r_s$ embedded between two larger regions 
of larger electron density. To have strictly 1d leads with negligible 
electron-electron interactions is certainly not realistic. If one uses 
semi-conductor heterostructures, to be outside the Luttinger-Tomonaga limit 
requires to take at least quasi-1d leads, if not 2d electron gases (2DEGs) 
of high enough densities. If the leads become two dimensional, the non local 
effect should have a faster decay ($1/L_c^2$, instead the $1/L_c$) with 
$\lambda_F/2$ oscillations. If the leads remain quasi-1d, the decay will be 
slower. 

  Eventually, let us mention transport measurements 
\cite{Topinka1,Topinka2,LeRoy} imaging coherent electron flow from a 
quantum point contact (QPC) where the non local effect induced by 
electron-electron interaction could play a role. They are made using a 2DEG 
created in a GaAs/AlGaAS heterostructure. A QPC cut the 2DEG in two parts, 
and a charged AFM tip can be scanned around the QPC. The QPC conductance $g$ 
is measured as a function of the AFM tip position. When $g$ takes a low 
value, the QPC is almost closed and the electron density is low around it, 
making very likely non negligible interaction effects. Let us note that 
such effects are believed to be crucial for the observed 0.7 ($2 e^2/h$) 
structure \cite{Thomas}. In the case of Refs. \cite{Topinka1,Topinka2,LeRoy}, 
the QPC is biased such that its conductance is on the first conductance 
plateaus ($g \approx 1, 2, 3$ in units of $2 e^2/h$). In that case, the 
QPC provides an interacting nano-system, while the external 
scatterer is given by the charged tip which creates a local depletion region 
in the 2DEG directly below it. It is observed that $g$ is changed when the 
tip is scanned around the QPC, the change $\delta g(L)$ being of order of a 
fraction of $2e^2/h$ and decaying \cite{Topinka1} as $1/L^2$ with the 
distance $L$ between the QPC and the tip. Moreover, a 2d plot of $g(L)$ as a 
function of the tip position shows \cite{Topinka1} fringes spaced by half 
the Fermi wave length $\lambda_F/2$. Therefore, the position-dependent 
conductance has exactly the behavior which one can expect if it is related 
to the mechanism described in this work, i. e. the behavior of Friedel 
oscillations in two dimensions. We leave to a further work a study of this 
2d set-up, for knowing if a quantitative description of the measured 
$\delta g(L)$ does not require to go beyond the non interacting electron 
picture, making necessary to take into account our non local effect, at 
least when $g$ is on the low conductance plateaus.     

\section{Acknowledgments}

We thank M.~Sanquer for drawing our attention to Refs. 
\cite{Topinka1,Topinka2,LeRoy} and D.~Weinmann for useful comments. 
The support of the network ``Fundamentals of nanoelectronics'' of the 
EU (contract MCRTN-CT-2003-504574) is gratefully acknowledged.

\end{document}